\documentclass[prb,twocolumn,showpacs,preprintnumbers,amsmath,amssymb,superscriptaddress]{revtex4}

\usepackage{graphicx}
\usepackage{dcolumn}
\usepackage{bm}
\usepackage{longtable}

\begin{document}

\preprint{APS/123-QED}

\title{Magnetically-Driven Ferroelectric Atomic Displacements in perovskite like YMnO$_3$
}

\author{D. Okuyama}%
\affiliation{Cross-Correlated Materials Research Group (CMRG), and Correlated Electron Research Group(CERG), 
RIKEN-ASI, Wako 351-0198, Japan}
\author{S. Ishiwata}%
\affiliation{Cross-Correlated Materials Research Group (CMRG), and Correlated Electron Research Group(CERG), 
RIKEN-ASI, Wako 351-0198, Japan}
\affiliation{Department of Applied Physics and Quantum-Phase Electronics Center (QPEC), University of Tokyo, Hongo, 
Tokyo 113-8656, Japan}
\author{Y. Takahashi}%
\affiliation{Multiferroics Project, ERATO, Japan Science and Technology Agency (JST), c/o  University of Tokyo, Hongo, Tokyo 113-8656, Japan}
\author{K. Yamauchi}%
\altaffiliation[Present address: ]{ISIR-SANKEN, Osaka Univ., Japan.}
\affiliation{Consiglio Nazionale del le Ricerche-Superconducting and Innovative materials and device (CNR-SPIN), 
67100, L'Aquila, Italy}
\author{S. Picozzi}%
\affiliation{Consiglio Nazionale del le Ricerche-Superconducting and Innovative materials and device (CNR-SPIN), 
67100, L'Aquila, Italy}
\author{K. Sugimoto}%
\affiliation{JASRI SPring-8, Hyogo 679-5198, Japan}
\author{H. Sakai}%
\affiliation{Cross-Correlated Materials Research Group (CMRG), and Correlated Electron Research Group(CERG), 
RIKEN-ASI, Wako 351-0198, Japan}
\author{M. Takata}%
\affiliation{RIKEN SPring-8 Center, Hyogo 679-5148, Japan}
\author{R. Shimano}%
\affiliation{Multiferroics Project, ERATO, Japan Science and Technology Agency (JST), c/o  University of Tokyo, Hongo, Tokyo 113-8656, Japan}
\affiliation{Department of Physics, University of Tokyo, Tokyo 113-8656, Japan}
\author{Y. Taguchi}
\affiliation{Cross-Correlated Materials Research Group (CMRG), and Correlated Electron Research Group(CERG), 
RIKEN-ASI, Wako 351-0198, Japan}
\author{T. Arima}%
\affiliation{RIKEN SPring-8 Center, Hyogo 679-5148, Japan}
\affiliation{Institute of Multidisciplinary Research for Advanced Materials, Tohoku University, Sendai 980-8577,
Japan}
\author{Y. Tokura}%
\affiliation{Cross-Correlated Materials Research Group (CMRG), and Correlated Electron Research Group(CERG), 
RIKEN-ASI, Wako 351-0198, Japan}
\affiliation{Department of Applied Physics and Quantum-Phase Electronics Center (QPEC), University of Tokyo, Hongo, 
Tokyo 113-8656, Japan}
\affiliation{Multiferroics Project, ERATO, Japan Science and Technology Agency (JST), c/o  University of Tokyo, Hongo, 
Tokyo 113-8656, Japan}

\date{\today}

\begin{abstract}
Magnetically-driven ferroelectric atomic displacements of the order of 10$^{-3} \rm{\AA}$ have been observed 
in orthorhombic (perovskite like) YMnO$_3$ by a single-crystal synchrotron x-ray diffraction. 
The refined polar structure shows the characteristic bond alternation driven by the exchange striction 
in staggered Mn-O-Mn arrays with $\uparrow\uparrow\downarrow\downarrow$ type ordering, 
giving rise to a spontaneous polarization along $a$-axis. 
First-principles calculations based on the Berry phase method as well as on the 
experimentally refined crystal structure can reproduce the observed polarization value. 
\end{abstract}

\maketitle


\section{Introduction}

A strong coupling among spin, charge, and lattice degrees of freedom is a key requirement for future spintronic devices. 
Magnetoelectric multiferroics is one such example satisfying this demand \cite{Cheong2007,Tokura2010}, 
as typified by the magnetic-order-induced polarization in orthorhombic (perovskite like) $R$MnO$_3$ ($R$: Tb and Dy) 
and Ni$_3$V$_2$O$_8$ with a cycloidal spin order \cite{Kimura2003,Kenzelmann2005,Lawes2005}. 
The relation between cycloidal spin structure and electric polarization has been extensively discussed in terms of 
spin current 
\cite{Katsura2005} or by the inverse Dzyaloshinskii-Moriya model 
\cite{Mostovoy2006,Sergienko2006a,Jia2006}, both of which consider 
the spin-orbit coupling as the relevant interaction. 
Subsequently, first-principles calculations using the density functional theory (DFT) were performed to explain 
the actual value of the spontaneous polarization in a quantitative manner\cite{Xiang2008,Malashevich2008}. 

In displacement-type ferroelectrics, 
where the electric polarization had been attributed to the displacements of the constituent ions in early days, 
an essential contribution of the quantum Berry phase of valence electrons 
was revealed by accurate structural analysis and first-principles calculation \cite{Cohen1992,Zhong1994,Kuroiwa2001}. 
Therefore, also in the multiferroic system, the collaboration between the theoretical calculation and the crystal structural 
analysis will give critical information to the microscopic origin of the ferroelectric polarization. 
Nevertheless, little has been clarified for the crystal structures of the ordered phases in multiferroics, primarily due to 
the extremely small lattice displacements driven by the magnetic order as reflected in the small polarizations, 
typically $<$1 $\mu$C/cm$^{2}$. 
For a typical multiferroic TbMnO$_3$, the intensities of the x-ray superlattice reflections with a wavenumber 2$q_m$ 
($q_m$: wavenumber of spin modulation), which appear depending on the type of spin helix \cite{Arima2007}, 
are four orders of magnitude smaller than that of the strong fundamental reflections. 
These are contrastive to the cases of conventional displacement-type ferroelectrics BaTiO$_3$ and PbTiO$_3$ 
with large polarization (several tens of $\mu$C/cm$^{2}$). 

Among the origins of spin-driven ferroelectricity, lattice striction induced by the symmetric spin exchange 
$\boldsymbol{S}_{i} \cdot \boldsymbol{S}_{j}$ has also been of great interest because of the potentially large polarization 
\cite{Sergienko2006b,Cheong2007}, which can be ascribed to the larger energy scale of the exchange interaction 
irrelevant to Dzyaloshinskii-Moriya interaction (spin-orbit coupling). 
Typical examples of exchange-striction-driven ferroelectrics are orthorhombic (perovskite like) $R$MnO$_3$ 
($R$= Y and Ho-Lu) showing the $\uparrow\uparrow\downarrow\downarrow$-type ($E$-type) spin order \cite{Munoz2002}, 
Ca$_3$(Co,Mn)$_2$O$_6$ \cite{Choi2008}, and GdFeO$_3$ \cite{Tokunaga2009}. 
As for the orthorhombic $R$MnO$_3$, the existence of the enhanced polarization of about 6 $\mu$C/cm$^{2}$ 
has been proposed in the $E$-type phase by the first-principles calculations based on the Berry phase method and 
theoretically optimized crystal structure \cite{Picozzi2007,Yamauchi2008}. 
Recently, the polarization in the $E$-type phase has been confirmed for polycrystalline samples to be much larger than 
those of the helimagnetic phase, but the value appears one order of magnitude smaller than the calculated value 
\cite{Lorenz2007,Ishiwata2010}. 
Such discrepancy should be solved by knowing the accurate crystal structure of the multiferroic state as well as 
by comparing the experimental values with the result of the first-principles band-theoretical calculation. 
In this context, the orthorhombic $R$MnO$_3$ with the $E$-type order is an ideal system to study the microscopic origin 
of the magnetically-driven ferroelectricity from the structural viewpoint; the advantages are that 
(i) the lattice displacements are expected to be relatively large because the polarization is particularly large among the 
magnetically-driven multiferroics, 
(ii) extensive reports regarding to the multiferroic properties have been accumulated for the series of the orthorhombic 
$R$MnO$_3$. 
In this paper, we report the observation of the atomic displacements in the ferroelectric phase of orthorhombic YMnO$_3$ 
by means of synchrotron x-ray diffraction for the single crystalline sample. 
The atomic displacements with alternating Mn-O-Mn bond angles were found to produce polarization along the $a$-axis 
(in the \textit{Pbnm} setting) in the $E$-type phase. 
By a first-principles calculation, we can quantitatively reproduce the experimentally observed polarization value for 
appropriate values of on-site Coulomb interaction $U$. 

\section{Experimental and computational procedures}

Single crystals of orthorhombic YMnO$_3$ were synthesized by a high-pressure treatment (1573 K and 5.5 GPa) 
on the precursor sample of hexagonal YMnO$_3$ with using a cubic-anvil-type high-pressure apparatus. 
Details of the crystal growth of orthorhombic YMnO$_3$ under high pressure will be reported elsewhere 
\cite{Ishiwata2011}. 
Although the as-prepared crystals with a dimension up to 500 $\mu$m were twinned with alternating the 
$a$- and $b$-axes, the typical single-domain size was as large as 100 $\sim$ 200 $\mu$m. 
While measurements of the polarization and the dielectric permittivity were carried out on a twinned crystal, 
the magnetic susceptibility, infrared reflectivity and synchrotron x-ray diffraction were measured for twin-free small crystals. 

For measurements of the dielectric permittivity and the electric polarization, gold electrodes were deposited on both 
faces of the rectangular crystal with dimensions of 200 $\times$ 240 $\times$ 200 $\mu$m$^3$. 
As a poling procedure, an electric field of up to 12 kV$/$cm was applied at 40 K, followed by cooling to 2 K, and 
then the electric field was turned off. 
Displacement current was measured with increasing temperature at a rate of 5 K$/$min, and was integrated as 
a function of time to obtain the polarization. 
For the $E\perp c$ measurement, electric field was perpendicular to the $c$-axis and slanted at an angle of 
approximately 33.5$^{\circ}$ from the $a$- ($b$-) axis for the major (minor) twin domain; 
this was confirmed by a polarizing microscope. 
(The actual twin domain boundaries are observed along [1 1 0] or [1 -1 0] directions.) 
The reduction of the polarization value due to this misorientation is corrected by 
multiplying the observed raw data by 1.2 ($=1/\cos{(2\pi\frac{33.5^{\circ}}{360^{\circ}})}$). 
Magnetic susceptibility was measured in a heating run at a magnetic field of 1 kOe with using a SQUID
(Superconducting Quantum Interference Device) magnetometer. 
A Fourier-transform spectrometer (Bruker IF66v/S) combined with a microscope (Hyperion) and a HgCdTe detector was 
used for infrared spectroscopy. 

Synchrotron x-ray diffraction experiments were performed on BL02B1 at SPring-8, Japan. 
The photon energy of the incident x-rays was tuned at 34.91 keV. 
X-ray beams were shaped into a square of 300$\times$300 $\mu$m$^2$ by a collimator, 
which was large enough to irradiate the whole sample. 
A single crystal was crushed into cubes with a typical dimension of about 50 $\mu$m. 
The absorption coefficient $\mu$ for 34.91 keV is calculated to be 47.4 cm$^{-1}$. 
Therefore, $\mu r$ was about 0.25, which certified that the absorption effect was small enough for 
the empirical absorption correction \cite{Higashi1995}. 
A large cylindrical imaging plate was utilized to measure diffracted intensities. 
The temperature was controlled by a helium gas stream cryostat. 
Rapid-Auto program (Rigaku Corp. and MSC.) was used to obtain an F-table. 
Diffraction data were collected up to the resolution of 0.2 \AA. 
SIR2004 \cite{Sir2004} and CRYSTALSTRUCTURE (Rigaku Corp. and MSC.) programs were used 
for analyzing the crystal structure from the F-table. 

As for first-principles calculations, we performed DFT simulations using the VASP code \cite{Kresse1996} within the 
GGA+$U$ \cite{Anisimov1997} formalism with various $U$ values. 
In addition, we used the Heyd-Scuseria-Ernzerhof (HSE) screened hybrid functional method, \cite{HSE} 
which mixes the exact non-local Fock exchange and the density-functional parametrized exchange. 
The HSE is known to improve the evaluation of the Jahn-Teller distortion, with respect to GGA+$U$ approaches
\cite{HSEslv}. 
The cut-off energy for the plane-wave expansion of the wavefunctions was set to 400 eV and a $\boldsymbol{k}$-point 
shell of (2, 3, 4) was used for the Brillouin zone integration according to Monkhorst-Pack special point mesh. 
Spin-orbit interaction is not taken into account. 

\section{Results and discussion}

\begin{figure}
\includegraphics*[width=70mm,clip]{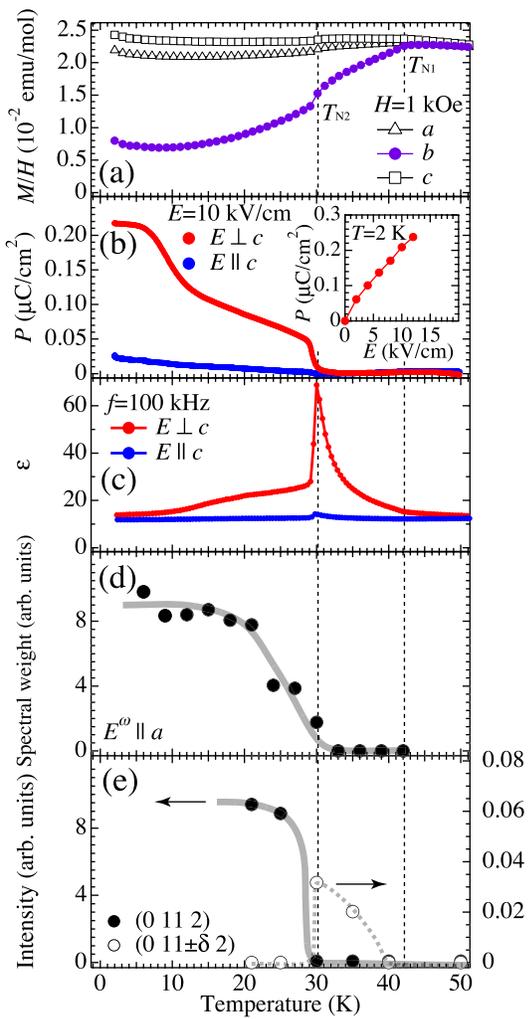}
\caption{\label{fig_1} (Color online) 
The temperature dependence of (a) magnetic susceptibility, (b) polarization ($P$), (c) electric permittivity ($\varepsilon$), 
(d) spectral weight of the activated 77 meV oxygen stretching phonon mode (see also Fig. \ref{fig_2} (b)), 
and (e) integrated intensity of synchrotron x-ray diffraction of a forbidden (0 11 2) reflection and satellites 
(0 11$\pm\delta$ 2). 
The inset to (b) shows the poling electric-field ($E$) dependence of the $P$ at 2 K. 
All the measurements were performed in heating process. 
The temperature dependence of polarization shows a step-like behavior which would probably arise from the 
phase coexistence occurring near the first-order phase boundary with the incommensurate cycloidal phase 
\cite{Ishiwata2010,Mochizuki2010,Takahashi2010}. 
The gray lines in (d) and (e) are the guide to the eyes. 
}
\end{figure}

\subsection{Magnetic and dielectric properties}

The temperature dependence of various physical properties of YMnO$_3$ measured on the single crystal 
is shown in Fig. \ref{fig_1}. 
Two kink-like anomalies are clearly observed at $T_{\mathrm{N1}}$=42 K and $T_{\mathrm{N2}}$=30 K, 
respectively, in magnetization along the $b$-axis, while they are barely discerned along the $a$- and $c$-axes 
(Fig. \ref{fig_1}(a)), corresponding to the magnetic phase transitions from the paramagnetic to the sinusoidal and then 
to the $E$-type structures with spins pointing almost along the $b$-axis 
\cite{Munoz2002,Lorenz2007,comment1}. 
In the $E$-type phase, the ferroelectric polarization perpendicular to the $c$-axis appears, as accompanied with a 
discontinuous jump of the dielectric permittivity $\varepsilon$ at $T_{\mathrm{N2}}$, as shown in Figs. \ref{fig_1}(b) and (c). 
In contrast, the anomaly of $\varepsilon$ along the $c$-axis is hardly discernible and the polarization along the 
$c$-axis is very small. 
These results indicate that ferroelectric polarization lies within the $ab$-plane in the $E$-type phase. 
While the polarization value at 2 K is not saturated at a poling field of 12 kV/cm as shown in the inset, 
the sample showed a breakdown for higher poling fields. 
Nevertheless, the genuine value of polarization at 2 K is expected to be at least 0.24 $\mu$C/cm$^{2}$. 

\subsection{Infrared reflectivity}

\begin{figure}
\includegraphics*[width=80mm,clip]{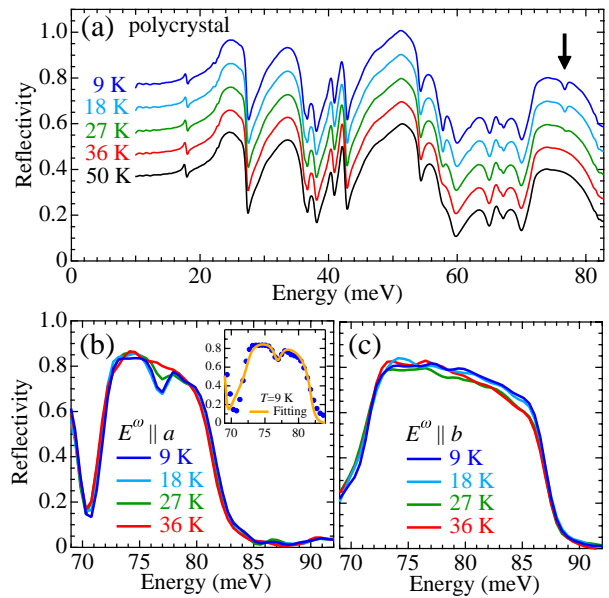}
\caption{\label{fig_2} (Color online) 
Infrared reflectivity spectra of the oxygen stretching phonon of (a) polycrystalline and single crystal samples 
for (b) $E^{\omega}\parallel a$ and (c) $E^{\omega}\parallel b$ both in the $E$-type and in the sinusoidal phase. 
Phonon anomaly is observed only for $E^{\omega}\parallel a$ configuration. 
A dip structure around 77 meV is attributed to an originally infrared-inactive but lowered-lattice-symmetry induced 
phonon mode. 
In the inset, the infrared data at 9 K and the fitting spectra with three Lorentz oscillators are shown. 
}
\end{figure}

Figure \ref{fig_2}(a) shows far-infrared reflectivity spectra of a polycrystalline sample. 
A lot of phonon modes are observed due to the orthorhombic distortion of the lattice 
from the cubic perovskite structure, where three infrared-active phonon modes, i. e. stretching, bending 
and external modes, are identified. 
A clear anomaly is observed around 77 meV below $T_{\mathrm{N2}}$ as indicated by an arrow. 
According to Ref.~\onlinecite{Kim2006}, a Raman-active mode at 77 meV gains the infrared activity due to the lowering 
of the symmetry in the ferroelectric phase. 
By using an as-grown single-crystal surface with a dimension of 50$\times$100 $\mu$m$^2$, 
the polarization of each phonon mode was clearly decomposed, as shown in Figs. \ref{fig_2}(b) and (c). 
The anomaly at 77 meV is observed only for $E^{\omega}\parallel a$ below $T_{\mathrm{N2}}$ indicating that 
the new optical phonon (oxygen stretching) mode is related to the loss of glide-plane normal to the $a$-axis. 
Therefore, the spectral weight of the mode can be a measure of the magnitude of the ferroelectric lattice displacement. 
To deduce the temperature dependence of the spectral weight, we fitted the reflectivity spectra above 65 meV 
for the $E^{\omega}\parallel a$ configuration by three Lorentz oscillators described as 
$S/(\omega^2-\omega_0^2-i\omega\gamma)$ (see the inset of Fig. \ref{fig_2}(b)). 
Here, $S$, $\omega_0$, and $\gamma$ are spectral weight, eigenfrequency, and inverse of lifetime of the oscillator, 
respectively. 
Because of the ambiguity of the absolute value of the reflectivity in this measurement, 
the scale factor for absolute reflectivity is another fitting parameter. 
The parameters ($S$, $\omega_0$, and $\gamma$) of two phonon modes were fixed to the value obtained by the 
fitting above $T_{\mathrm{N2}}$. 
The parameters of the Raman-active phonon mode were obtained for each temperature below $T_{\mathrm{N2}}$. 
The intensity of this mode increases with decreasing temperature below $T_{\mathrm{N2}}$ and 
does not vary much below around 20 K, as shown in Fig. \ref{fig_1}(d), indicating the saturation of the lattice displacement. 

\subsection{Synchrotron x-ray diffraction and crystal structure analysis}

\begin{figure}
\includegraphics*[width=70mm,clip]{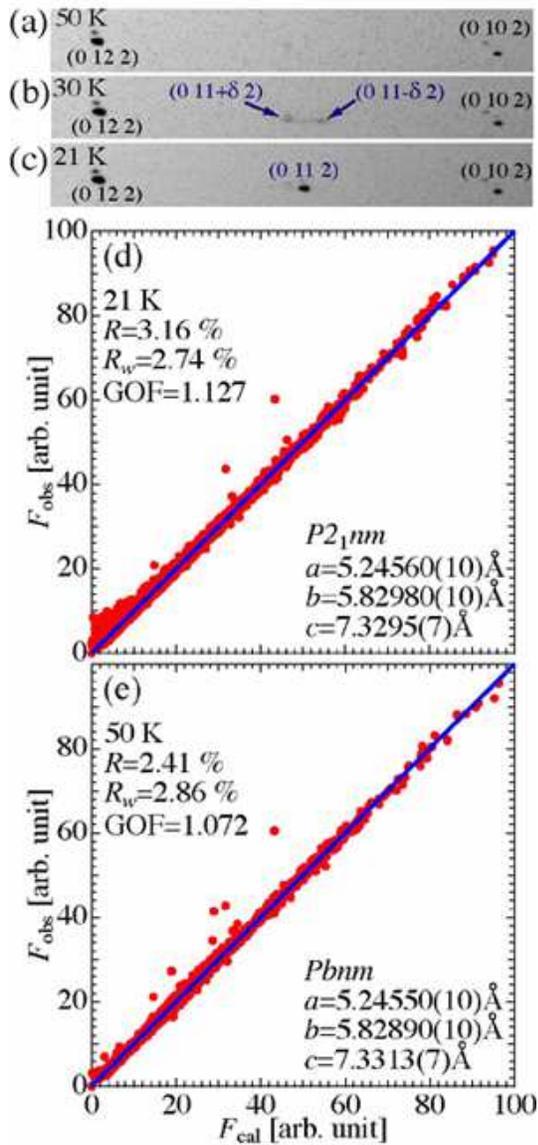}
\caption{\label{fig_3} (Color online) 
Synchrotron x-ray oscillation photographs in the (a) paramagnetic (50 K), (b) sinusoidal (30 K), and (c) $E$-type 
phase (21 K). 
At 21 K($T<T_{\mathrm{N2}}$), a (0 11 2) reflection forbidden in the \textit{Pbnm} symmetry is clearly observed, 
indicating the disappearance of the $b$-glide operation. 
At 30 K ($T_{\mathrm{N2}}<T<T_{\mathrm{N1}}$), (0 11 2) reflection splits along the $b^{*}$-axis, resulting in two 
incommensurate (0 11$\pm\delta$ 2) superlattice spots with $\delta$=0.08. 
Comparison between observed ($F_{\mathrm{obs}}$) and calculated ($F_{\mathrm{cal}}$) structure factor 
(d) at 21 K in the ferroelectric phase and (e) at 50 K in the paraelectric phase. 
}
\end{figure}

\begin{figure}
\includegraphics*[width=80mm,clip]{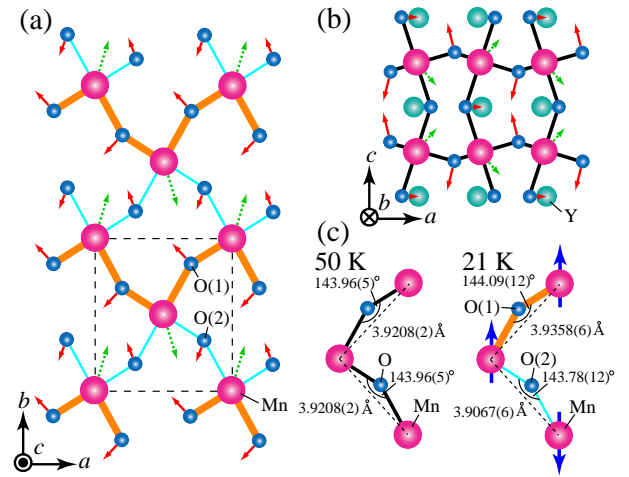}
\caption{\label{fig_4} (Color online) 
Schematics of the ferroelectric atomic displacements projected onto (a) the $ab$-plane and (b) the $ac$-plane 
in the ferroelectric $E$-type phase. 
Dotted and solid arrows indicate the directions and the relative atomic displacements of Mn and O, 
respectively. 
Note that the displacements of O ions are doubled for the purpose of clarify. 
A unit cell is indicated by a broken line. 
(c) Changes in the Mn-O-Mn bond angles and the distance between the neighboring Mn ions in the $ab$-plane. 
In the ferroelectric phase (21 K), the $E$-type spin configuration is represented by large arrows. 
Here, we assumed the ferromagnetic (antiferromagnetic) spin arrangement for Mn-O-Mn bond with larger (smaller) 
bond angle. 
}
\end{figure}

\begin{table*}
\caption{\label{table_1}
Structure parameters of orthorhombic YMnO$_3$ at 21 K in the ferroelectric phase 
(Space group \textit{P2$_1$nm} (No. 31)). 
The lattice parameters are $a$=5.24560(10) \AA, $b$=5.82980(10) \AA, and $c$=7.3295(7) \AA, respectively. 
In the tables, $x$, $y$, and $z$ are the fractional coordinates. 
Anisotropic atomic displacement parameters are represented as $U_{11}$, $U_{22}$, $U_{33}$, $U_{12}$, $U_{13}$, 
and $U_{23}$. 
}
\begin{ruledtabular}
\begin{tabular}{ccccccc}
 &site&$x$&$y$&$z$\\
\hline
Y1 & 2$a$ & 0.61103(7) & 0.83569(3) & 1/2\\
Y2 & 2$a$ & 0.64624(7) & 0.66579(3) & 0\\
Mn & 4$b$ & 0.12942(10) & 0.74832(4) & 0.24935(5)\\
O1 & 4$b$ & 0.4273(3)& 0.57846(18) & 0.3044(2)\\
O2 & 2$a$ & 0.0168(3) & 0.7118(2) & 1/2\\
O3 & 2$a$ & 0.2409(4) & 0.7868(2) & 0\\
O4 & 4$b$ & 0.8291(3) & 0.92158(18) & 0.1971(2)\\
\end{tabular}
\begin{tabular}{ccccccc}
 &$U_{11}$ (\AA$^2$)&$U_{22}$ (\AA$^2$)&$U_{33}$ (\AA$^2$)&$U_{12}$ (\AA$^2$)&$U_{13}$ (\AA$^2$)&
 $U_{23}$ (\AA$^2$)\\
\hline
Y1 & 0.00094(3) & 0.00169(2) & 0.00156(4) & -0.00020(3) & 0 & 0\\
Y2 & 0.00201(4) & 0.00169(3) & 0.00142(4) & 0.00007(3) & 0 & 0\\
Mn & 0.001510(10) & 0.002200(10) & 0.001450(10) & 0.000270(10) & 0.000040(10) & 0.000170(10)\\
O1 & 0.0031(3) & 0.0033(2) & 0.0039(3) & 0.0009(2) & -0.0002(2) & -0.00016(16)\\
O2 & 0.0024(4) & 0.0045(3) & 0.0024(3) & 0.0001(3) & 0 & 0\\
O3 & 0.0039(4) & 0.0033(2) & 0.0024(3) & 0.0008(3) & 0 & 0\\
O4 & 0.0028(2) & 0.0038(2) & 0.0030(2) & 0.0000(2) & -0.0004(2) & -0.00037(15)\\
\end{tabular}
\end{ruledtabular}
\end{table*}

\begin{table*}
\caption{\label{table_2}
Structure parameters of orthorhombic YMnO$_3$ at 50 K in the paraelectric phase 
(Space group \textit{Pbnm} (No. 62)). 
The lattice parameters are $a$=5.24550(10) \AA, $b$=5.82890(10) \AA, $c$=7.3313(7) \AA. 
}
\begin{ruledtabular}
\begin{tabular}{ccccccc}
 &site&$x$&$y$&$z$\\
\hline
Y & 4$c$ & 0.017610(10) & 0.084830(10) & 1/4\\
Mn & 4$b$ & 1/2 & 0 & 0\\
O1 & 4$c$ & 0.61224(8) & 0.96229(7) & 1/4\\
O2 & 8$d$ & 0.29892(5) & 0.32844(5) & 0.05366(3)\\
\end{tabular}
\begin{tabular}{ccccccc}
 &$U_{11}$ (\AA$^2$)&$U_{22}$ (\AA$^2$)&$U_{33}$ (\AA$^2$)&$U_{12}$ (\AA$^2$)&$U_{13}$ (\AA$^2$)&
 $U_{23}$ (\AA$^2$)\\
\hline
Y & 0.001800(10) & 0.001680(10) & 0.001700(10) & 0.000090(10) & 0 & 0\\
Mn & 0.001800(10) & 0.002210(10) & 0.001660(10) & -0.000250(10) & -0.000020(10) & 0.000230(10)\\
O1 & 0.00349(7) & 0.00389(7) & 0.00259(5) & -0.00060(6) & 0 & 0\\
O2 & 0.00325(5) & 0.00361(5) & 0.00357(4) & 0.00049(4) & -0.00031(3) & -0.00029(3)\\
\end{tabular}
\end{ruledtabular}
\end{table*}

Synchrotron x-ray oscillation photographs for the respective phases are displayed in Figs. \ref{fig_3}(a), (b), and (c). 
At 50 K above $T_{\mathrm{N1}}$, only fundamental Bragg reflections for \textit{Pbnm} symmetry are observed 
(Fig. \ref{fig_3}(a)). 
In the sinusoidal phase, two incommensurate reflections of (0 11$\pm\delta$ 2) appear, as indicated by arrows 
in Fig. \ref{fig_3}(b) ($\delta$=0.08 at 30 K). 
The intensity and $\delta$-value increases and decreases with decreasing temperature, respectively. 
As shown in Fig. \ref{fig_3}(c), at 21 K in the $E$-type phase, the two incommensurate reflections merge into a 
reflection (0 11 2), which is forbidden in the original \textit{Pbnm} symmetry \cite{comment1}. 
As shown in Fig. \ref{fig_1}(e), the integrated intensity of the additional peaks shows a second-order phase-transition 
like temperature dependence in the sinusoidal phase with the incommensurate modulation, while exhibiting 
a first-order-like sudden variation in intensity and $\delta$ value ($\delta$=0.08 to $\delta$=0) upon 
the transition from the sinusoidal to the $E$-type phase. 

By using the data sets of synchrotron x-ray diffraction, we performed crystal structure analyses in the $E$-type 
(i.e. ferroelectric) and the paramagnetic (i.e. paraelectric) phases, respectively. 
The comparisons between observed and calculated structure factors are shown in Figs. \ref{fig_3}(d) and (e). 
At 21 K in the ferroelectric phase, 37305 reflections were observed, and 10500 of them were independent. 
54 parameters were used for the refinement. 
The lattice parameters $a$, $b$, and $c$ are 5.24560(10) \AA, 5.82980(10) \AA, and 7.3295(7) \AA, respectively. 
Because the forbidden reflections of (0 $k$ $l$): $k$=$odd$ with \textit{Pbnm} setting are observed, 
the plausible space group is \textit{P2$_1$nm} (No. 31), a maximal non-isomorphic subgroup of \textit{Pbnm}. 
The disappearance of $b$-glide reflection allows a uniform atomic displacement along the $a$-axis. 
In the structure analysis, the flack parameter \cite{Flack1988} is estimated as 0.503(18), indicating that 
the volume fractions of two kinds of ferroelectric domains are nearly equal. 
The reliability factors are $R$= 3.16\%, $R_{w}$=2.74\%, GOF(Goodness of fit)=1.127. 
At 50 K in the paraelectric phase, 15409 reflections were observed, and 5668 of them were independent. 
29 parameters were used for the refinement. 
The lattice parameters are as follows: $a$=5.24550(10) \AA, $b$=5.82890(10) \AA, $c$=7.3313(7) \AA. 
The space group is \textit{Pbnm} (No. 62). 
The reliability factors are $R$= 2.41\%, $R_{w}$=2.86\%, GOF=1.072. 
The crystal parameters at 21 K in the ferroelectric phase and at 50 K in the paraelectric phase are 
summarized in Tables \ref{table_1} and \ref{table_2}, respectively. 

The crystal structure analyses indicated the atomic displacements of Mn and O ions within the $ab$-plane and 
the $ac$-plane in the ferroelectric phase as shown by arrows in Figs. \ref{fig_4}(a) and (b), respectively. 
Here, the positions of Y ions are fixed. 
The displacements of Mn ions along the $a$-axis are observed to be uniform, while those along the $b$- and 
$c$-axes are staggered. 
The shifts of O ions along the $a$-axis are also uniform but opposite to those of Mn ions. 
The components of the O displacement along the $b$- and $c$-axes entirely cancel out, similarly to those of Mn ions. 
These uniform atomic displacements of Mn and O ions along the $a$-axis and the associated variation of electron 
wave functions generate the ferroelectric polarization parallel to the $a$-axis. 
The observed atomic displacements along the $a$-axis are as small as 0.0041(7) \AA\ for Mn ion and 0.0020(16) 
and 0.0030(16) \AA\ for two inequivalent O sites, respectively. 
These values are one or two orders of magnitude smaller than those of conventional displacement-type ferroelectrics, 
BaTiO$_3$ and PbTiO$_3$ \cite{Kuroiwa2001}. 
From the point charge model with nominal ionic charges (Y; 3+, Mn; 3+, O; 2-) and these values of atomic 
displacement along the $a$-axis, ferroelectric polarization value is calculated to be approximately 0.5 $\mu$C/cm$^{2}$, 
which is in the same order as the experimentally observed value. 
Changes of the Mn-O-Mn bond angles and the distance between neighboring Mn ions are detailed in Fig. \ref{fig_4}(c). 
At 50 K, there is only one kind of Mn-O-Mn bond with an angle of 143.96(5)$^{\circ}$ in basal $ab$-plane. 
By contrast, at 21 K, there are two inequivalent bonds of Mn-O(1)-Mn and Mn-O(2)-Mn in $ab$-plane, whose bond angles, 
144.09(12)$^{\circ}$ and 143.78(12)$^{\circ}$, are slightly larger and smaller than that of Mn-O-Mn in the paraelectric 
phase, respectively. 
In other words, an alternating bond structure emerges along the $b$-axis in the ferroelectric phase, as shown 
in Fig. \ref{fig_4}(a). 

The mechanism of the polarization generation has been proposed to be a symmetric-exchange striction that provides 
the gain of magnetic exchange energy. 
So as to minimize the total nearest-neighbor exchange energy mediated by the local ferromagnetic double-exchange 
interaction, the bond angles for ferromagnetic $\uparrow\uparrow$ (antiferromagnetic $\uparrow\downarrow$) 
Mn spin arrangement become large (small), 
because the transfer integral of $e_{g}$ electron increases (decreases) for the larger (smaller) Mn-O-Mn angle 
\cite{Yamauchi2008,Mochizuki2010}. 
The result of the present x-ray structure analysis is in accord with the basic idea of the exchange-striction model. 

\subsection{Theoretical calculation}

\begin{figure}
\includegraphics*[width=70mm,clip]{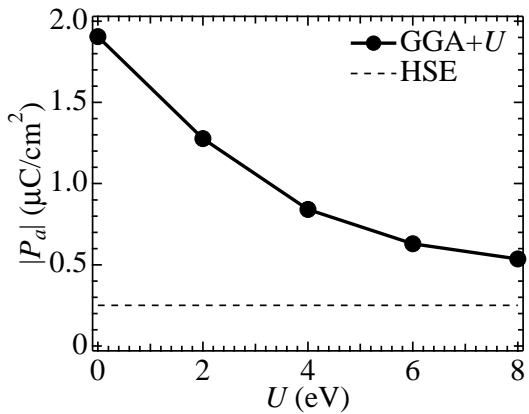}
\caption{\label{fig_5} 
The absolute value of calculated ferroelectric polarization in the $E$-type phase by using the Berry phase technique. 
Solid circles represent the results obtained from the GGA+$U$ method and the horizontal dashed line corresponds 
to the result from HSE calculations. 
The simulations were performed on the basis of the experimentally refined atomic coordinates of the ferroelectric-state 
crystal (at 21 K). 
}
\end{figure}

For a more detailed discussion, we performed first-principles simulations of the YMnO$_3$ electronic structure. 
In this paper, since the optimization of the geometry - in terms of atomic positions - is a rather delicate issue 
(see comment below), we stick to the experimentally refined crystal structure, as now available from the detection of 
atomic displacements of 10$^{-3}$ \AA\ order. 
The calculated ferroelectric polarization $P_a$ using the Berry phase method \cite{King1993} for the experimental 
structure points to $-a$ direction, and the absolute value $|P_a|$ is plotted versus the Hubbard parameter 
$U$ in Fig. \ref{fig_5}. 
The $|P_a|$ value decreases from 1.8 to 0.5 $\mu$C/cm$^{2}$ when increasing $U$ from 0 to 8 eV. 
The reduction of $|P_a|$ is mainly due to the reduction of hopping, in turn linked to the reduction of the $e_g$-related 
double-exchange interaction. 
The polarization value is rather strongly affected by the $U$ value, which is, unfortunately, experimentally not known for 
YMnO$_3$. 
It is however experimentally reported that $U$ is in the range 5$\sim$7.5 eV in the case of a similar compound, 
LaMnO$_3$ \cite{Saitoh1995,Park1996}, so that the $|P_a|$ value can be considered to be 0.5$\sim$0.8 $\mu$C/cm$^{2}$. 
In order to overcome the problems related to the arbitrary choice of the Hubbard $U$ parameter, we performed 
hybrid-functional (HSE) calculations and obtained an even smaller value of 0.25 $\mu$C/cm$^{2}$. 
As for the comparison with experiments, we remark that theoretical $|P_a|$ values are larger than the observed values 
$\sim$0.08 $\mu$C/cm$^{2}$ at 21 K and $\sim$0.24 $\mu$C/cm$^{2}$ at 2 K. 
One possible origin for this discrepancy is the insufficiency of the poling field in the experiment (see the inset of 
Fig. \ref{fig_1} (b)). 
The corrected value of the ferroelectric polarization observed for the polycrystalline sample is approximately 0.5 
$\mu$C/cm$^{2}$ near 2 K\cite{Ishiwata2010} in agreement with the theoretically predicted order of magnitude. 
This improvement, compared to earlier studies \cite{Picozzi2007,Yamauchi2008} is due to the use of experimental 
structure parameters; the bond angle difference is experimentally determined to be 0.3$^{\circ}$, while previous bare-DFT 
calculations were based on 3$^{\circ}$ of bond angle difference, obtained from the theoretical optimized crystal structure. 
Despite some advances in the comparison between theory and experiments, several issues  remain to be solved, 
in particular related to the geometry optimization of the crystal structure. 
Density functional theory, while capturing Mn-O-Mn bond-angles and Mn-Mn bond-lengths alternating features, 
indicated \cite{Picozzi2007,Yamauchi2008} opposite direction of Mn and O displacements along the $a$-axis 
as compared with the experimentally refined structure. 
While the magnitude of the relaxation effects depends on the $U$ value, the sign is the same for every $U$, 
as well as when using hybrid-functionals approaches. 
These results are probably related to the overestimate of $p$-$d$ hybridization for (more itinerant) $e_g$ electrons and/or 
to the underestimate of the interaction between (more localized) $t_{2g}$ spins. 
It is possible that magnetostrictive effects related to $t_{2g}$ states, expected to produce opposite trends for the Mn-O-Mn 
angles (i.e. larger angles for antiparallel Mn spins), are not properly described within all  DFT, DFT+$U$ or HSE approaches. 
Finally, the assumption of a fully collinear spin structure might also play a role in the theory-experiments comparison. 
We therefore hope that our work will stimulate further theoretical activities aimed at quantitatively reproducing the correct 
experimental crystal structure and magnetostrictive effects in orthorhombic manganites. 

\section{Summary}

In summary, we have performed measurements of the magnetic susceptibility, electric polarization, electric permitivity, 
infrared reflectivity, and synchrotron x-ray diffraction to investigate the multiferroic nature of orthorhombic YMnO$_3$ 
by using the single crystal. 
The results indicate that the ferroelectric phase transition with the polarization along the $a$-axis takes place 
concurrently with the E-type magnetic phase transition. 
In the ferroelectric phase, the phonon anomaly and forbidden reflection are respectively observed in the infrared reflectivity 
and synchrotron x-ray diffraction, indicating the magnetically-induced lattice distortion; 
the space group changes to \textit{P2$_1$nm} polar structure. 
The atomic displacements of the order of 10$^{-3} \rm{\AA}$ are estimated from the crystal structure analysis 
in the ferroelectric phase, indicating the formation of the bond alternating structure of Mn-O-Mn bond in $ab$-plane. 
The feature is qualitatively consistent with theoretical predictions of the exchange-striction model as a microscopic 
mechanism for generating the electric polarization. 
Further, by the collaboration between the crystal structure analysis and first-principles calculations, 
we quantitatively explain the observed macroscopic polarization value with the input of experimental information of 
the microscopic atomic displacement. 
This work may pave a path to understanding the microscopic origins of the magnetically-driven 
ferroelectricity from the structural point of view with the support of the first-principles calculations, 
which would facilitate further structural studies of other magnetically-driven ferroelectrics 
with small lattice displacements. 

\begin{acknowledgments}
The authors are grateful to Y. Tokunaga for furuitful discussions. 
The synchrotron x-ray diffraction experiment was performed at SPring-8 with approval of the JASRI 
(Proposal Numbers 2009B1304 and 2010A1795). 
The theoretical research was supported by the European Community, FP7/ERC Grant Agreement no. 203523. 
This work was in part supported by Funding Program for World-Leading Innovative R\&D on Science 
and Technology (FIRST Program) from the JSPS 
and Grants-in-Aid for scientific research (No. 19052001 and 20046017) from MEXT, Japan. 
\end{acknowledgments}


\end{document}